# Insights into the room temperature magnetism of ZnO/Co$_3$O$_4$ mixtures.

*M. S. Martín-González*

Instituto de Microelectrónica de Madrid, CSIC. Tres Cantos Madrid, Spain

*J. F. Fernández, F. Rubio-Marcos, I. Lorite,*

Instituto de Cerámica y Vidrio, CSIC, 28049 Madrid, Spain

*J. L. Costa-Krämer,*

Tate lab of Physics, University of Minnesota, Minneapolis, Minnesota 55455, USA

On leave from: Instituto de Microelectrónica de Madrid, CSIC. Tres Cantos Madrid, Spain

*A. Quesada,*

Instituto de Magnetismo Aplicado (RENFE-UCM-CSIC), Las Rozas Madrid, Spain

*M. A. Bañares, J. L. G. Fierro*

Instituto de Catálisis y Petroleoquímica, CSIC 28049 Madrid, Spain.

The origin of room temperature ferromagnetic like behavior (RT-FM) in ZnO-based diluted magnetic semiconductors (DMS) is still an unclear question. The present work concentrates on the appearance of Room temperature magnetic moments in just mixed ZnO/Co$_3$O$_4$ mixtures without thermal treatment. In this study is shown that the magnetism seems to be related to surface reduction of the Co$_3$O$_4$ nanoparticles. In which, an antiferromagnetic Co$_3$O$_4$ nanoparticle (core) is surrounded by CoO-like shell. This singular superficial magnetism has also been found in others mixtures with semiconductors like TiO$_2$ and insulators like Al$_2$O$_3$.

**Introduction**

The discovery and understanding of room temperature ferromagnetic semiconductors based on Dilute Magnetic Semiconductors (DMS) is a grand challenge in material science, since they would be excellent candidates to be used in the next generation of spintronic devices.[1-2] According to the theoretical predictions,[3] doping certain semiconductors, such as ZnO, with a few transition metal atoms like Mn or Co would lead to this type of behavior. Although, the mixtures of ZnO with Mn and Co oxides exhibit room temperature ferromagnetism (RTFM) extensive doubts persist about the origin and interaction between magnetic and semiconducting properties.[4] Nowadays, there are three different approaches in literature devoted to determine the origin of magnetism and its relation with the semiconducting properties. a) A first one, related to the doping of the transition metals like Mn or Co inside the ZnO lattice [5,6] where conduction electrons become spin-polarized, so these materials are, in fact, magnetic semiconductors and for that useful for spintronic applications. b) A second approach, in which the magnetism is explained to be due not only to the presence of magnetic ions, but also to the presence of ZnO oxygen defects.[7-9] And, c) a third one, in which new discoveries point toward a different direction: these materials are not DMS and the magnetism is shown to be due either to a secondary phases[10] or to an interfacial double-exchange mechanism at the diffusion front of Zn cations inside the $MnO_2$ particles, as we reported[11] for the case of $ZnO:MnO_2$

Recently, just mixed $ZnO:Co_3O_4$ nanopowders without further thermal treatment show a well-defined hysteresis loops at 300K,[12] which is indicative of a robust high temperature magnetic behavior. As compared with pure ZnO and $Co_3O_4$ which showed no RTFM. As it was observed, the magnetic signals of the $ZnO:Co_3O_4$ mixtures were low enough to associate their origin to magnetic impurities; however, $M_s$ decreases when increasing the Co concentration in the samples and with the thermal treatment at 400ºC and became non-magnetic for thermal treatment up to 800ºC, thus other possible mechanisms could not be discarded. Due to the importance of the appearance of RT magnetic behavior just by mixing ZnO with $Co_3O_4$, this study focused in-depth on this point. In addition, a study of the $Co_3O_4$ mixtures with other oxides with insulating properties demonstrates that this phenomenon is not only related with ZnO.

**Experimental Section**

ZnO:Co$_3$O$_4$ samples with a ratio of: a) 99% ZnO, 1% Co$_3$O$_4$; b) 95% ZnO, 5% Co$_3$O$_4$; and c) 75% ZnO, 25% Co$_3$O$_4$ (named ZC1, ZC5 and ZC25, respectively from now on) were fabricated following the low temperature procedure described previously:[12-13] high purity ( >99.9% Aldrich) ZnO with average particle size of 500 nm and Co$_3$O$_4$ nanoparticles raw powders were used for sample preparation. Also, non-mixed samples were prepared in the same conditions for comparison purposes. In addition 1% Co$_3$O$_4$ mixtures with TiO$_2$ and Al$_2$O$_3$ were also prepared following the above procedure. The average particle size was selected to be similar to ZnO.

The structural characterizations were carried out with a Siemens D5000 X-ray diffractometer. High-Resolution Transmission Electron Microscopy (HRTEM) was performed in a JEOL JEM 3000F. Charge to mass ratio is measured with a "Faraday cage" consisted in a double metal enclosure that were isolated electrically and the induced charge measured of the dried samples was measured with a Keithley High Resistance Electrometer model 6517A. Photoelectron spectra (XPS) were acquired with a VG ESCALAB 200R spectrometer provided with an AlK$_\alpha$ X-ray source (1486.6 eV). Raman spectra were collected on a Renishaw Micro-Raman System 1000. The samples were excited with the 514 nm Ar line. Magnetic properties were measured in a high temperature VSM above RT.

**Results and Discussion**

In all of the samples mixed at RT, the presence of Co$_3$O$_4$ and ZnO was determined by XRD. No displacement in the X-ray diffraction peak positions was detected and no additional peaks related with other Co/Zn oxides were found. So, at this scale a reaction of components were not identified by XRD. Nevertheless, these mixtures present room temperature magnetic behavior while pure ZnO and Co$_3$O$_4$ do not.[12]. In order to understand this change in the magnetic properties, a more in detail study was performed.

The ZnO and the Co$_3$O$_4$ particles agglomerates as observed by TEM, figure 1a, and they stick together even after 1.5 h of ultrasonic treatment. This stickiness can be explained by the different superficial charge of the starting powders, which is – 0.44 nC/g for ZnO and + 0.22 nC/g for $Co_3O_4$, as measured in a Faraday cage.

Pure ZnO and $Co_3O_4$ nanoparticles are found by HRTEM. The images show ordered structures. The planar distances of the ZnO particles correspond to pure wurtzite structure although EDS find a small amount of Co (around 1%). This apparent doping is not likely and appears related to the nearness of cobalt nanoparticles that would contribute to the X-rays signal recorded. The $Co_3O_4$ nanoparticles present a cubic spinel structure cubic with $Co^{2+}$ and $Co^{3+}$ located at tetrahedral and octahedral sites, respectively, which belongs to the space group (*Fd3m*) and cell parameters of 8.0837 Å, figure 1b, in which, $Co^{2+}$ is in tetrahedral and $Co^{3+}$ is in octahedral coordination. In case of the cobalt oxide nanoparticles, the EDS composition of the particles is around 99% Co and 1% Zn that reflects the above-mentioned detection limit. Nevertheless for these particles, if the doping occurs the cell parameters should not change since $Co_3O_4$ and $ZnCo_2O_4$ are isostructural with a very close unit cell. The presence of Co metal or other cobalt oxides can be excluded. Therefore, the magnetic properties of the mixture are attributed to the $Co_3O_4$ or ZnO particles, rather than to ferromagnetic impurity or the exchange coupling between Co metal and $Co_3O_4$.

Since the previous techniques did not give an explanation to the magnetic response observed a further study of the structural changes in $Co_3O_4$ and ZnO particles was performed by XPS, see figure 2. Pure $Co_3O_4$ spectra is characterized by a Co $2p_{3/2}$ peak at 780.2 eV a shake-up satellite peak at 789.3 eV, a Co $2p_{1/2}$ peak at 795.6 eV and the satellite at 804.8 eV. The separation among the spin-orbit component (15.4 eV), the satellites shape and its positions correspond with pure $Co_3O_4$. [14] In the case of ZnO: $Co_3O_4$ mixtures, no big changes are observed either in the $2p_{3/2}$, $2p_{1/2}$ peak positions or in the separation between them. This means that particles are essentially pure $Co_3O_4$ and no in-diffusion of Zn into de $Co_3O_4$ particles is detected. It should also be noted that the binding energies of these peaks are largely independent of the Zn:Co ratio, which also suggests that the local electronic structure of the cobalt oxides does not change with the compositions and emphasized that no Zn doping is produced.

However, important changes are detected in the satellite peaks. They became more intense and appear at lower binding energies (785 and 803.5 eV). This position is representative of octahedral coordinated $Co^{2+}$, cations similar to those present in the CoO structure. The fact that the positions of the spin-orbit components remain unchanged and the satellites become more similar to CoO structure has been assigned before to $Co_3O_4$ with CoO like-structure at the surface.[15] This means that some kind of reduction at the surface level is taking place just by mixing $Co_3O_4$ and ZnO in an attrition miller in water with zirconia balls. This agrees with the HRTEM and the XRD results, since the bulk of the particles is $Co_3O_4$ and only a few $Co^{3+}$ ions (in octahedral coordination) of the surface are reduced to $Co^{2+}$. It is also interesting to emphasize that the observed FM in not induced by metallic Co-clusters since no metallic cobalt was detected by HRTEM or XPS. In the case of the ZnO particles, Zn $2p_{3/2}$ core level peak appears at 1021.7 eV suggesting that cobalt has not in-diffuse into the ZnO lattice so Co is not dopping the ZnO structure.

The superficial reduction of the $Co_3O_4$ nanoparticles was confirmed by Raman spectroscopy. A displacement of the bands is observed after mixing $Co_3O_4$ with ZnO when compare with pure $Co_3O_4$, see figure 3. This displacement can be interpreted as a partial reduction of the $Co_3O_4$ nanoparticles to CoO. This displacement is more noticeable for ZC1, in which $Co_3O_4$ particles see more ZnO during the milling process. Then, it can be concluded that the role of ZnO is to promote this superficial reduction.

Based on the previous results, it could be reasonable to propose that the room temperature ferromagnetic-like behavior is due to a core–shell model in which an antiferromagnetic nanoparticle ($Co_3O_4$) is surrounded by a CoO-like shell. Thus, the anomalous magnetic properties must be related to an interaction between the shell spins (in which $Co^{2+}$ is in octahedral coordination) and core spins (in which $Co^{2+}$ is in tetrahedral coordination). The finding supports that the origin of this FM is not related to the doping required for the Diluted Magnetic Semiconductor approach but to a superficial effect. This magnetism is different from the one that we report previously since there is not heat treatment or diffusion front or new phase formation.[11] But, it a "new magnetism" ascribed to the particle surface.

In order to determine if this phenomenon is only due to the interaction of $Co_3O_4$ with ZnO or it is more general, the same preparation procedure was performed by mixing 1% $Co_3O_4$ with other oxides - with semiconducting and insulating properties- such as $TiO_2$ and $Al_2O_3$ (figure 2). The starting $TiO_2$ powder exhibits a room temperature magnetic behavior with an Ms of 1 x $10^{-4}$ emu/$g_{TiO2}$, probably due to impurities of the starting powders. However, a strong increase of >60% was detected in the $TiO_2$/$Co_3O_4$ mixture. Moreover, room temperature magnetism is also found when $Co_3O_4$ is mixed with an insulator with pure diamagnetic signal, like $Al_2O_3$ (figure 2). This stresses the relevance of this surface magnetism, which appears in mixtures with paramagnetic and diamagnetic behaviour, but also with semiconducting and insulating properties. The role of these materials seems to be the reduction of the $Co_3O_4$ surface, although it can not be rule out that these materials also could align the uncompensated surface spins of $Co_3O_4$ though an interfacial induction, since they are sticked close together.

In conclusion, it have been established that the presence of room temperature magnetic properties in the just mixed ZnO/$Co_3O_4$ samples is not related to a DMS behavior. Appears to be due to core-shell model in which an antiferromagnetic $Co_3O_4$ core (in which $Co^{2+}$ is in tetrahedral coordination and $Co^{3+}$ in octahedral coordination) is surrounded by a CoO-like shell (in which $Co^{2+}$ is in tetrahedrical coordination). Moreover, this magnetism is not only exclusively related to the interaction between $Co_3O_4$ and ZnO particles since it is observed with other oxides such $TiO_2$ and $Al_2O_3$, with semiconducting and insulating properties. Thus, we are dealing with an effect related with the nanoscale that lead to a room temperature ferromagnetic like behavior not discussed previously.

Acknowledgements

This work was supported by CYCYT MAT2007-66845-C02-01, CSIC 2006-50F0122 and CSIC 2007-50I015

Figure Captions:

Figure 1. HRTEM of ZnO:$Co_3O_4$ mixture. A) low magnification shown the interaction between ZnO and $Co_3O_4$ particles and B) high magnification of a $Co_3O_4$ particle on the [001] zone axis. (Inset) resultant FFT corresponding with a Spinel structure.

Figure 2. Photoemission spectroscopy of Co 2p core level of: pure $Co_3O_4$ and ZC1, ZC5 and ZC25. The satellite peak is displaced in all the just mixed samples.

Figure 3. Raman spectra comparison of the different mixtures ZC1, ZC5 and ZC25 as milled.

Figure 4. M versus H curves for different mixtures of 1% wt $Co_3O_4$ with oxides like $TiO_2$ and $Al_2O_3$ at room temperature (the paramagnetic contribution was subtracted).

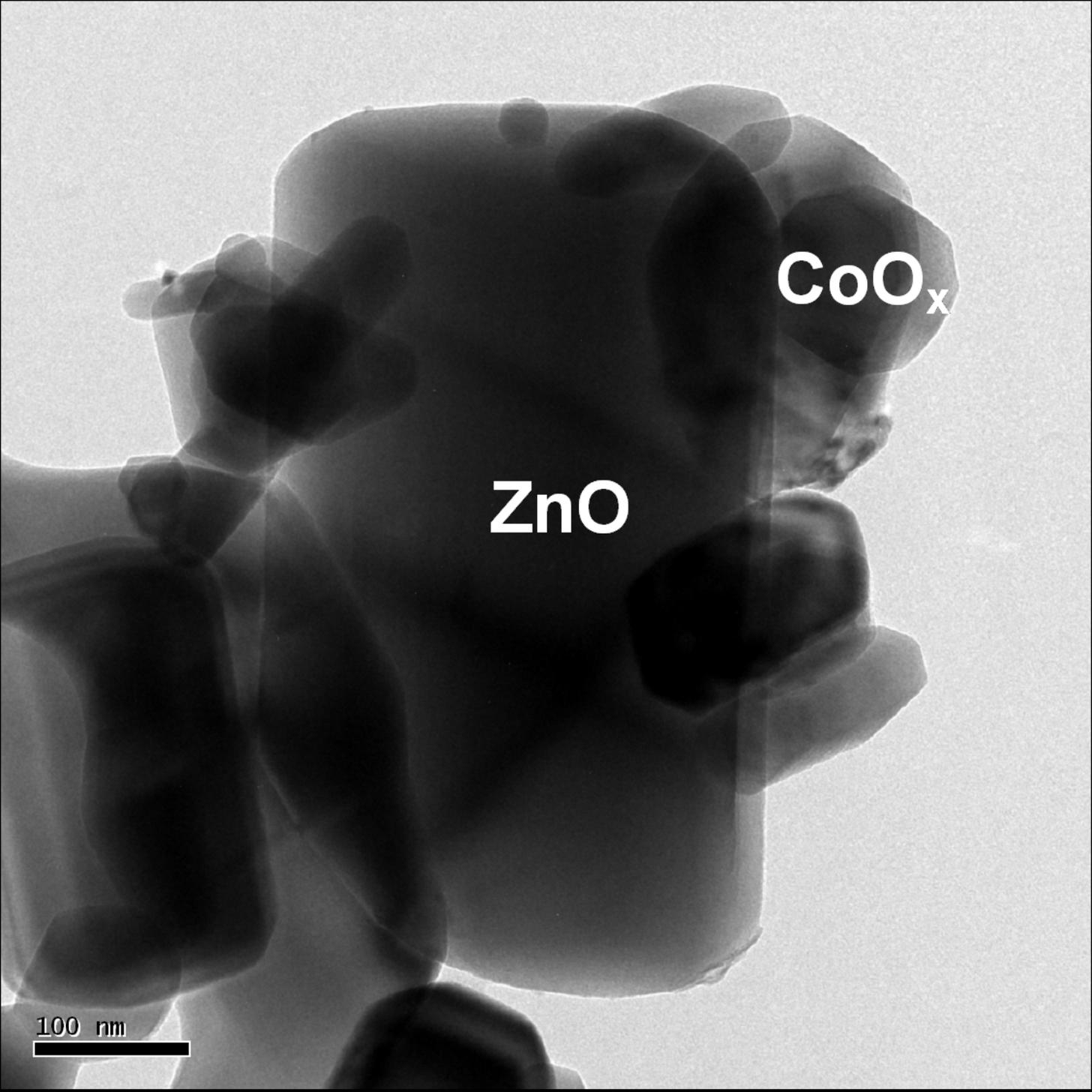

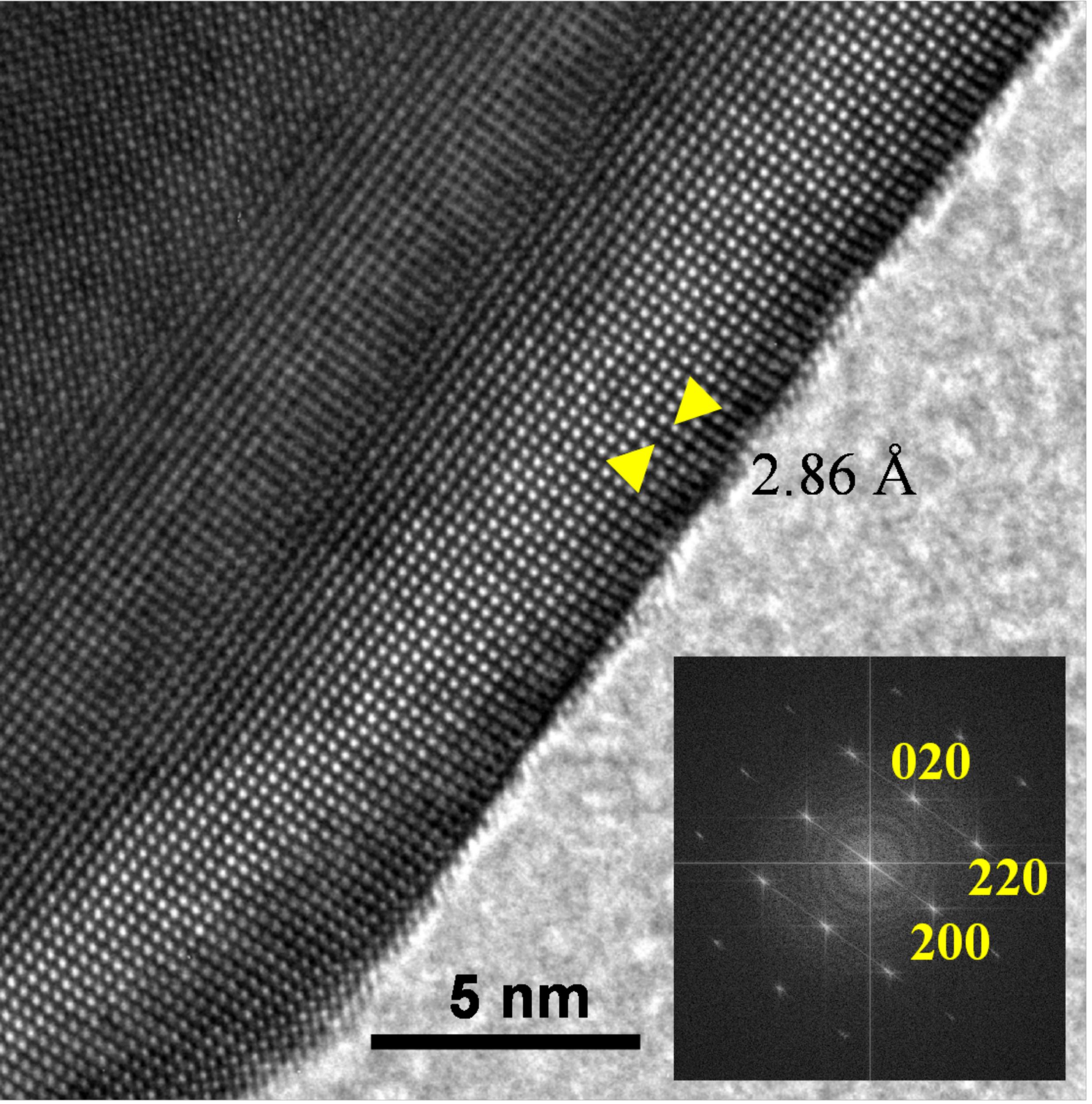

2.86 Å

020

220

200

5 nm

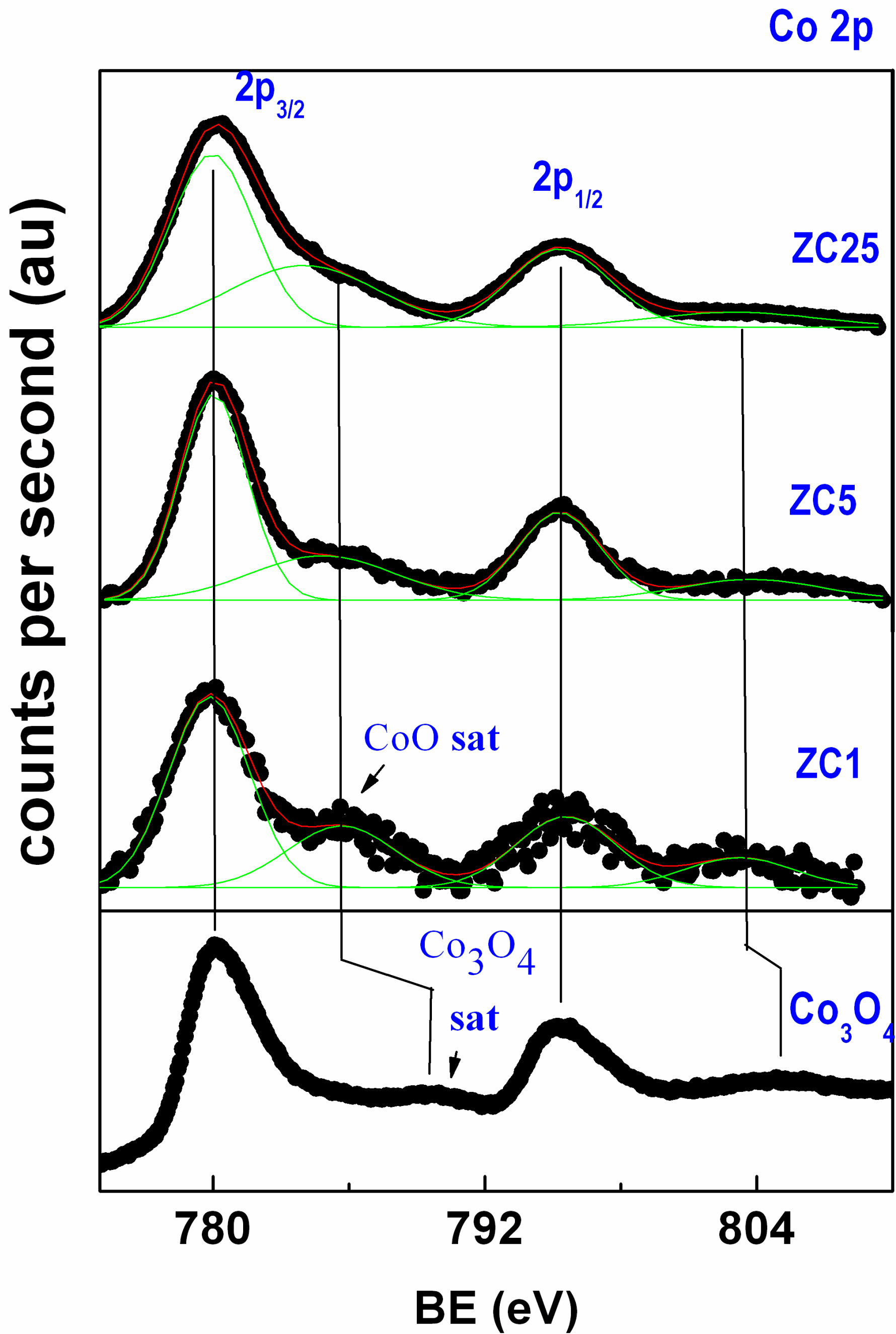

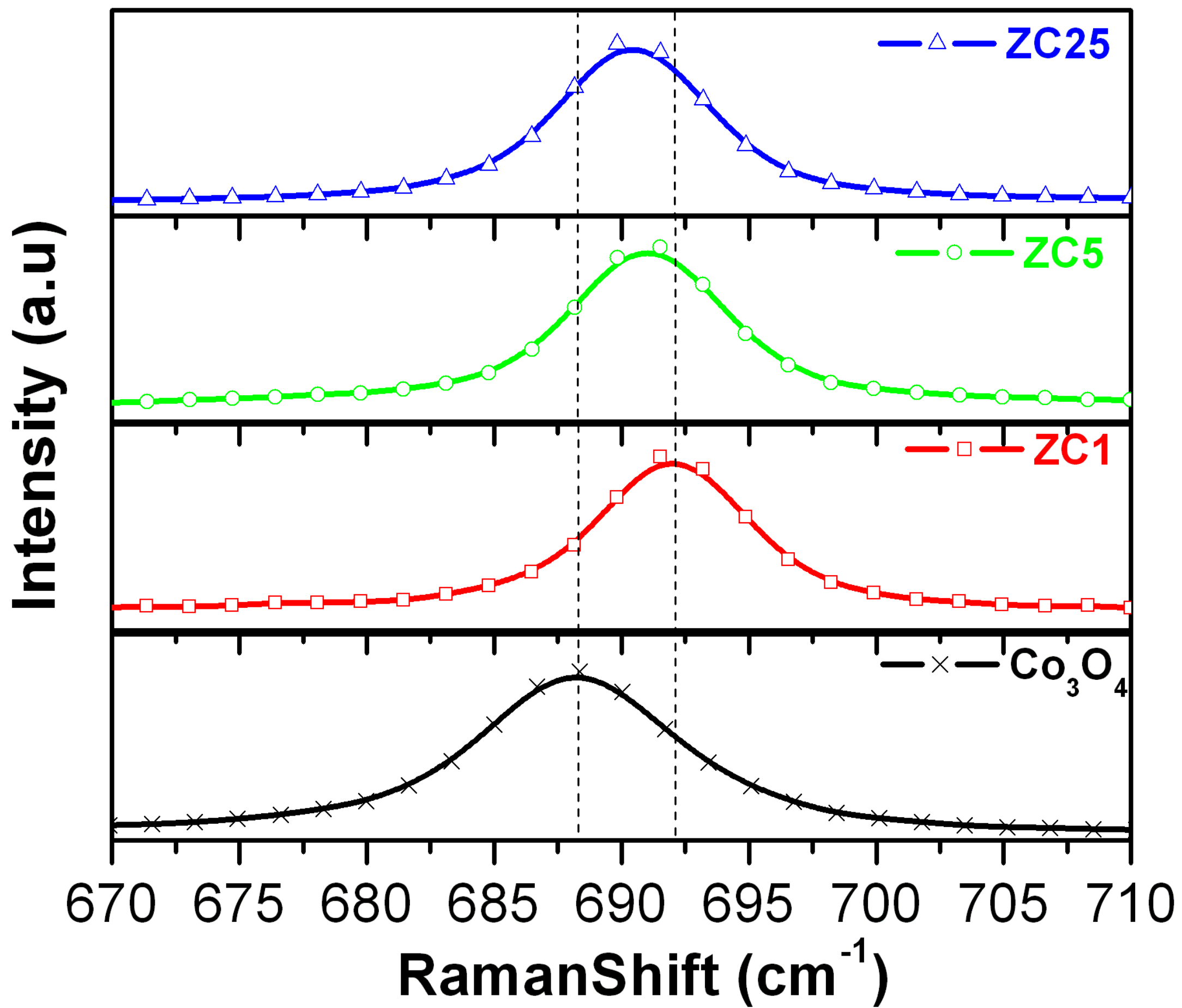

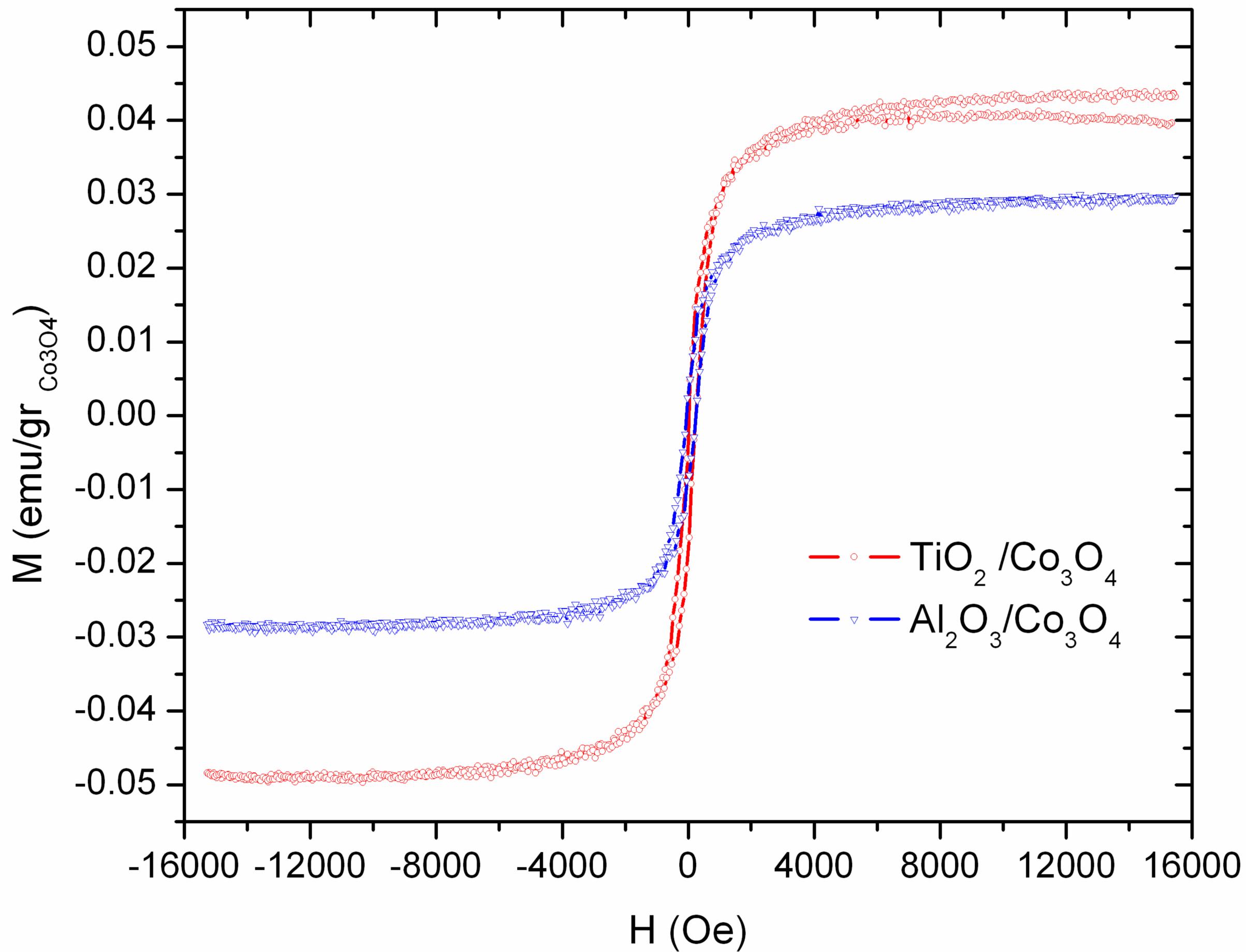